\documentclass[dvips]{article}
\usepackage{icrctc07}

\title{The VERITAS standard data analysis}
\shorttitle{The VERITAS standard data analysis}
\authors{M.~K.~Daniel$^{1}$ for the VERITAS collaboration$^{2}$}
\shortauthors{M.~K.~Daniel et al.}
\afiliations{$^1$School of Physics and Astronomy, University of Leeds, Leeds, LS2 9JT. U.K. \\
             $^2$for full author list see G. Maier, ``VERITAS: Status and Latest Results", these proceedings}
\email{mkd@ast.leeds.ac.uk}

\abstract{
VERITAS is an array of Imaging Atmospheric Cherenkov Telescopes designed for 
very high energy gamma ray (E$>$100\,GeV) observations of astrophysical sources. 
The experiment began its scientific observation program in the 2006/2007 
observing season. We describe here the analysis chain for reducing the data, 
reconstructing the direction and energy of incident gamma rays and the 
rejection of background cosmic rays.
}

\begin{document}
\maketitle

\section{Introduction}
% Should I describe data file size?
The VERITAS instrument, described in more detail in \cite{VERITAS}, consists of 
an array of 4 telescopes each equipped with 499 pixel cameras \cite{FPI} 
read out by 500\,MS/s flash-ADCs (fADC) \cite{DACQ}. The traces are recorded 
into a custom data format before being archived for distribution to the collaboration 
and further analysis. 
%Diagnostic plots are made from the data in these
%files in addition to information stored in a database recording the individual 
%telescope and array trigger rates, deadtime and sky clarity as measured by an 
%infra-red radiometer attached to one of the telescopes and other useful 
%quantities. The diagnostic information is made available through a
%web page for people to make data quality checks prior to starting an indepth
%analysis of any runs.
Diagnostic plots -- such as individual telescope and array trigger rates,
deadtime stabilitity, atmospheric clarity, etc -- are generated from these files
and database stored information. These plots are made available through a web 
page for data integrity checks prior to an analysis being attempted.
A separate web page listing plots showing the long term
history of the camera pixels (trigger participation, voltage information, etc)
in order to trace camera health is also available. % repository

A number of packages have been developed for the analysis of VERITAS data. 
Having multiple, independent packages available guards against any systematic 
errors being introduced into the analysis chain. Every analysis of a putative
source must then have an independent secondary analysis in order to confirm 
any results.
This paper describes the chain appropriate to the eventdisplay \cite{Holder06} 
%(used in the publication of prototype and single telescope VERITAS data)
and the VEGAS \cite{VEGAS} 
%(developed as a flexible analysis framework for array level reconstruction) 
packages. There are others, such as GrISU \cite{M87}
and the quicklook %ref?
package (used for online analysis when an observation is in progress) which
show consistent results to the ones given here. 

\section{Data Analysis}
%Once a run has been 
%deemed as acceptable 
%for further analysis 
The data analysis chain has three distinct elements: firstly the pixel
data is calibrated to get accurate charge information; after calibration
the pixel data is parameterised to give telescope level images, which
in turn can be combined to reconstruct the geometrical properties of the shower;
finally an estimate of the background level of counts is made in order to
determine if there are any statistically significant excesses (i.e.~sources) in
the field of view (fov). Once a source has been identified a data analysis can
be further extended to look for variability in emission with a light curve
and a spectrum calculated to try and distinguish between different models of the
emission process; or if there is no significant signal an upper limit can be
placed on the source emission. Specific details of the higher level analysis
process are left to the papers describing the analysis of specific objects.
%neither of those steps are gone into here because we ain't showing any spectra in the preliminaries :-(

\subsection{Calibration}
The technical details of obtaining calibration information are covered in
\cite{calibration}. Laser runs are used to calculate the relative gain between
the pixels and the timing offsets to counteract any time spread due to path
length differences (e.g. differing cable lengths, etc) for each channel. 
Pedestal events are injected at 1\,Hz during an observation run to give
an estimate of both the expected voltage offset for each fADC trace and an 
estimate of the night sky background (NSB) and electronics noise from the 
variance of that value. 
VEGAS allows the pedestal variance to be calculated as a function of time into
the run, with a default window of 3 minutes being the average time that
a star will take to pass through a pixel's fov. Pixels with values
outside of an expected range for the calibration parameters are then flagged
as bad and removed from further analysis.
%: those with too large a value are most likely too noisy to extract a reliable signal; those with too low a value are probably not working. 
The charge values for these problem tubes are set to 0 and 
typically this will apply to $\leq5\%$ of channels across the array (depending
on the number of bright stars in the fov). 
%though the exact number will depend on how many photomultiplier tubes needed to be turned off due to bright stars in the fov.

The amount of charge in a good trace is then calculated by summing the
samples for a given window size. In VEGAS this is defaulted to a 14\,ns
integration window that starts from a position calculated by averaging 
the pulse arrival time (defined as the half-height point on the pulse leading 
edge) of a representative number of events. 
In eventdisplay a two pass scheme is used. The first pass has a wide 20\,ns 
window that begins at a fixed position in order to calculate the integrated 
charge and the pulse arrival time. The second pass tightens the integration 
window to 10\,ns in order to increase the signal to noise ratio, placing the 
start of the integration according to the calculated time gradient across the 
image after it has been parameterised, as discussed in the next section.

\subsection{Parameterisation and reconstruction}
In the process of cleaning an image for parameterisation, pixels which produce 
an integrated charge greater than 5 times their pedestal standard deviation
(picture pixel) and any pixel 
that is adjacent to these higher threshold pixels and having 2.5 times 
their own standard deviation (boundary pixel) are automatically assigned to 
the image. 
Isolated picture pixels, i.e.~those with no other picture or boundary pixels 
neighbouring them, and all other pixels then have their charges set to 0. 
The resulting shower image is parameterised with a second moment analysis and 
for eventdisplay the time gradient across that image is found for the second
pass of the charge integration. 

%Quality cuts are size, ntubes, distance, angle between image axis.
%threshold after quality cuts? - in sims paper presumably.
%a table of these cuts?
%efficiency of reconstruction?
%analysis cut values?
%Q-factors?

The image data from the individual telescopes is then tested against standard 
quality criteria for the number of tubes present in an image is $\geq5$
(to ensure a robust image axis), 
the minimum amount of charge in an image is $size > 400\,$digital counts 
(providing run-to-run and NSB fluctation stability)
%bring things above the hardware trigger threshold
%(to bring it sufficiently above the hardware trigger threshold to ensure 
%good run to run consistency and reduce any systematics due to variations in the
%NSB) 
and the angular distance in the camera is $0.05\leq dist \leq1.3$ degrees 
(avoiding bias due to image truncation).
%(to guard against any truncation of the shower image biasing reconstruction). 
If a sufficient number of telescope images for an event pass the quality 
cuts the analysis proceeds to extract stereoscopic information like the 
direction on the sky and the shower core position on the ground for each event. 
%then come the sims.
The image information can then be compared to lookup tables of the parameters 
for simulated gamma rays as a function of image size, impact distance
to the shower core and the zenith angle of observation to achieve a greater
rejection power of the uneven light distribution of cosmic ray events to 
that of the smooth compact gamma ray ones. 
Details of simulations of the VERITAS array can be found in \cite{sims}. 
For VEGAS the default method follows the prescription given in \cite{Daum97} 
where a mean scaled parameter (MSP) is given by
%\begin{equation}\label{msw}
%MSW = \frac{1}{N_{tel}}\sum_{k=1}^{N_{tel}}\frac{w_{i}}{\langle w \rangle_{k}^{ij}}
%\end{equation}
\begin{equation}\label{msw}
MSP = \frac{1}{N_{tel}}\sum_{i=1}^{N_{tel}}\frac{p_{i}}{\bar{p}_{sim}(\theta, size, r)}
\end{equation}
where $p_{i}$ corresponds to the parameter in question (width or length) for
telescope $i$ and $\bar{p}_{sim}(\theta, size, r)$ is the mean value for 
simulations of a given $size$ at a given impact distance ($r$) and zenith angle
$\theta$. Cuts based on the mean scaled parameters for VEGAS are 
$0.05\leq MSW \leq1.02$, $0.05\leq MSL \leq1.15$ (where $W$ stands for width and $L$
for length) and have a quality factor of $Q \sim 1.8$.
In eventdisplay the normalized width method described in \cite{Henric06} is 
utilised to reduce the impact of outliers in the distributions.
\begin{equation}\label{mscw}
MSCP = \frac{1}{N_{tel}}\sum_{i=1}^{N_{tel}}\frac{p_{i} - \tilde{p}_{sim}(\theta, size, r)}{\sigma_{90}}
\end{equation}
where $\tilde{p}_{sim}(\theta, size, r)$ is the median value; 
and $\sigma_{90}$ is the width of the distribution for 90\% of the events. 
The cuts are $-1.2\leq MSCW \leq 0.5$, $-1.2\leq MSCL \leq 0.5$ with a quality 
factor of this form of cut is $Q \sim 4$. The values of the cuts given here
should be considered standard for all results presented, but are not necessarily
considered as fully optimised.

\subsection{Background estimation}
%table of cuts and quality factors?
%cuts? Currently being optimised -- sensitive to simulation accuracy
%No sims paper for KASCADE sims used in VEGAS!
There is no set prescription for background estimation that can cope with every
circumstance or source morphology. As such the best way to keep systematic 
uncertainties under control is by applying several methods to the same data set 
and comparing the results, a detailed description of commonly used background 
estimation models can be found in \cite{Berge07}.

%what about on/off? -- good for extended source or multiple sources in the fov
%The simplest method is to pair on and off source observations at equal zenith
%angle -- the background is assumed to be equal between the runs and the
%difference provides an estimate of the gamma-ray signal.

\begin{figure}
\begin{center}
\includegraphics [width=0.45\textwidth]{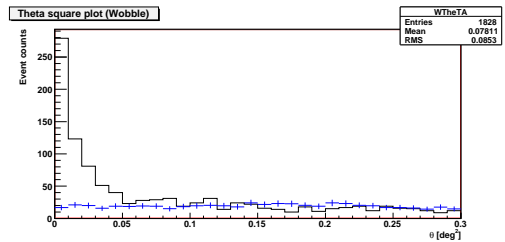}
\end{center}
\caption{
$\theta^{2}$ plot for 4 runs of Crab nebula observations taken with a wobble
offset of 0.5 degrees is given by the line. 
The background estimate, made from 5 circular regions of 0.22 degree 
diameter each, is given by the crosses.
}
\label{Wobble}
\end{figure}

%Wobble
Wobble mode observations \cite{Daum97} are favoured for point-like (or limited
extension) sources since a background estimate can be derived from the events
recorded during a run, allowing more on-source time and systematic effects 
in the background estimation due to variation in weather or performance changes 
effectively cancel out.
%A large fraction of the observations are taken for a point source that is at a 
%known position in the field of view. By taking an observation with the source 
%position offset by a fixed distance from the centre of the camera (though
%alternated in RA or Dec between observation runs to avoid camera biases) a
%background estimate can be derived from the events recorded during the run 
%simultaneously gathered data
%and systematic effects in the background estimation due to variation in weather 
%or performance changes effectively cancel out. 
%The background estimate is made from the number of events in a region located 
%on the opposite side of the camera to the assumed source position, i.e. the 
%region reflected around the camera centre. 
The background estimate is made from a region reflected from the source
position around the camera centre (the telescope pointing position). In order 
to gain a better estimate of the number of background events multiple 
background regions can be used, each region set in a ring the same offset 
distance from the camera centre, but still avoiding the region around the 
suspected source position to ensure that poorly reconstructed gamma ray events 
do not bias the background estimate. Point sources show up as an excess of 
reconstructed shower directions close to the assumed source position. 
Figure~\ref{Wobble} shows just such a plot for 5 background regions on 4 good 
quality Crab nebula $0.5^\circ$ wobble offset observations of 20 minute 
duration each. The spread of events closely matches that of a point 
source with an angular resolution of $\sim 0.14^\circ$ \cite{sims}.
%A plot of the square of the angular distance of the reconstructed gamma ray 
%source direction to the assumed source position, called a $\theta^{2}$ plot 
%and shown in figure~\ref{Wobble} for a selection of 4 good quality Crab nebula 
%observations of 20 minute duration each, will show a point gamma ray source as 
%an excess of events to small values of $\theta^{2}$. The Crab nebula is 
%essentially a point source to the telescopes and the fit of the point spread 
%function from simulations to the Crab nebula shows good agreement for an 
%angular resolution of $\sim 0.14^\circ$ (the angular resolution is energy 
%dependent, since higher energy showers give better images to reconstruct the 
%source position estimate to). 
After a cut for a point like source of $\theta^{2}<0.025$ for three telescope
data ($\theta^{2}<0.035$ for two telescope data) the quality factor for all cuts 
is $Q \sim 24$.

\begin{figure}
\begin{center}
\includegraphics [height=0.3\textheight]{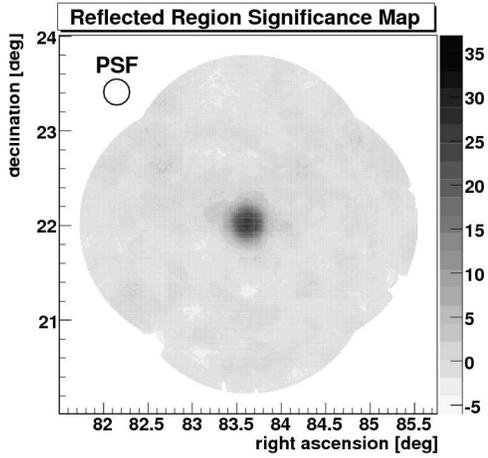}
\end{center}
\caption{\small{
2-d sky map of significances with the reflected region model for the same
observations as figure~\ref{Wobble}.}
}
\label{ReflectedRegion}
\end{figure}

%extend wobble to reflected region 2-d sky map.
The reflected region methodology for wobble mode observations can essentially be
applied to any part of the field of view displaced from the observation
position. This allows a 2-d map of events to be built up for the fov,
figure~\ref{ReflectedRegion} shows the 2-d distribution of significances for the
same Crab dataset using this method. A 2-d map like this will show up emission 
offset from the speculated source position, sources of extended emission and a 
modest number of multiple sources in the same fov that a 1-d analysis would 
miss.

%RBM as an alternative
An alternative 2-d mapping method is the ring background model.
In this model a ring (in celestial co-ordinates) around a trial source position 
is used to give the background estimate.
%Give radius of ring
%describe acceptance correction?
%radial symmetry
Since the ring covers areas with different offsets from that of the trial source
position an acceptance correction function must be used in the normalisation for 
each position on the ring. Any part of a ring that crosses an assumed source 
position is also excluded from the background estimate.

\begin{figure}
\begin{center}
\includegraphics [height=0.3\textheight]{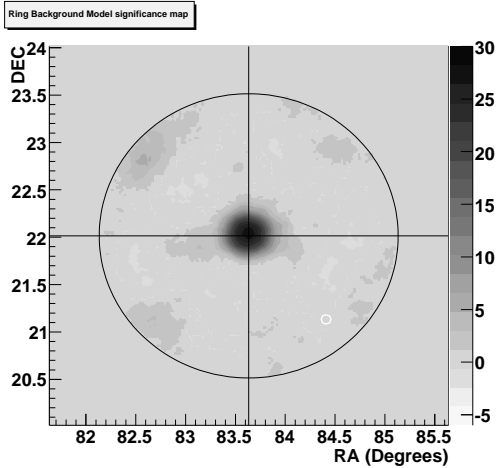}
\end{center}
\caption{\small{
2-d sky map of significances with the ring background model for the same
observations as figure~\ref{Wobble}. The white circle corresponds to the
position of zeta tau.}
}
\label{RBM}
\end{figure}

\begin{figure}
\begin{center}
\includegraphics [height=0.25\textheight]{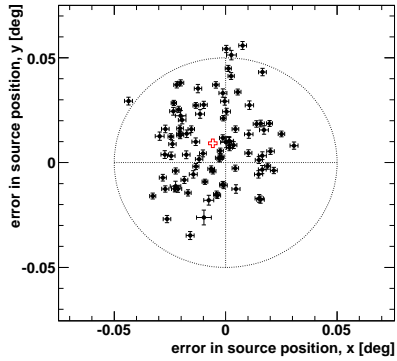}
\end{center}
\caption{\small{
Reconstruction of the source position for a large number of Crab nebula runs
(black squares). The red open cross is the overall reconstructed source position.}
}
\label{sourcePosXY}
\end{figure}

The source location accuracy can be seen in figure~\ref{sourcePosXY} which shows
the results for the fit of a 2-d Gaussian to the excess source counts. It can
be seen that a source position can be accurately reconstructed to less than
$0.05^\circ$ for each run.

%angular resolution -- sims paper?
%core reconstruction -- sims paper?
%source position location.
%significance as a function of offset?
%energy resolution -- no spectra in draft papers!

%table too large to fit!
%\begin{table}
%\begin{center}
%\begin{tabular}{c|c|c|c|c}
%package & pixels in image & image size & angular distance in camera & minimum angle
%between image axes \\
% & & [digital counts] & [degrees] & [degrees] \\
%\hline
%eventdisplay & $\geq 4$ & $\geq 400$ & $<1.3$ & $> 5$ \\
%VEGAS & $\geq 5$ & $\geq 400$ & $<1.3$ & $> 10$ \\
%\end{tabular}
%\caption{\label{cleaning}Quality cuts applied before stereo reconstruction.}
%\end{center}
%\end{table}

%is this table relevant?
%\begin{table}
%\begin{center}
%\begin{tabular}{c|c|c}
%package & cut & Q-factor \\
%\hline
%eventdisplay & -1.2<mscw<0.5 & 1.5 \\
%VEGAS & MSW < peak + 0.09 & 1.77 \\
%\end{tabular}
%\caption{\label{analysis}Analysis cuts.}
%\end{center}
%\end{table}

\section{Summary}
%significance/sqrt(hour)?
The standard VERITAS data analysis chain and the use of independent software 
packages to keep systematic errors under control has been described. 
The use of different background estimation models allows hypothesis testing of 
different source morphologies and searching for unidentified sources within the 
field of view. The results from two different analysis packages and three 
different background estimation proceedures provide consistent results of 
$\sim 30\,\sigma/\sqrt{\mathrm hour}$ for 3 telescopes on the Crab nebula for a 
wobble offset of $0.5^\circ$.

\small{
\subsection*{Acknowledgements}
This research is supported by grants from 
the U.~S.~Department of Energy, 
the U.~S.~National Science Foundation, 
the Smithsonian Institution, 
by NSERC in Canada, 
by PPARC in the U.~K.
and by Science Foundation Ireland 

\small{
\nocite{Berge07}
\nocite{Daum97}
\nocite{calibration}
\nocite{DACQ}
\nocite{Holder06}
\nocite{Henric06}
\nocite{FPI}
\nocite{VERITAS}
\nocite{VEGAS}
%This is the reference to .bib file (Without .bib!)
\bibliography{icrc0283}

\begin{thebibliography}{10}

\bibitem{Berge07}
D.~{Berge}, S.~{Funk}, and J.~{Hinton}.
\newblock {\em A\&A}, 466:1219--1229, 2007.

\bibitem{M87}
P.~{Colin} and {et al.}
\newblock {Observations of M87 with VERITAS}.
\newblock In {\em these proceedings}, 2007.

\bibitem{Daum97}
A.~{Daum} and {et al.}
\newblock {\em Astroparticle Physics}, 8, 1997.

\bibitem{calibration}
D.~{Hanna} and {et al.}
\newblock {Calibration techniques for VERITAS}.
\newblock In {\em these proceedings}, 2007.

\bibitem{DACQ}
E.~{Hays} and {et al.}
\newblock {VERITAS Data Acquisition}.
\newblock In {\em these proceedings}, 2007.

\bibitem{Holder06}
J.~{Holder} and {et al.}
\newblock {\em Astroparticle Physics}, 25:391--401, 2006.

\bibitem{Henric06}
H.~{Krawczynski} and {et al.}
\newblock {\em Astroparticle Physics}, 25:380--390, 2006.

\bibitem{sims}
G.~{Maier} and {et al.}
\newblock {Simulations studies of VERITAS}.
\newblock In {\em these proceedings}, 2007.

\bibitem{VERITAS}
G.~{Maier} and {et al.}
\newblock {VERITAS: Status and Latest Results}.
\newblock In {\em these proceedings}, 2007.

\bibitem{FPI}
T.~{Nagai} and {et al.}
\newblock {Focal plane instrumentation of VERITAS}.
\newblock In {\em these proceedings}, 2007.

\bibitem{VEGAS}
J.~{Quinn} and {et al.}
\newblock {VEGAS , the VERITAS Gamma-ray Analysis Suite}.
\newblock In {\em these proceedings}, 2007.

\end{thebibliography}
%This in the bibtex style, is ok.
\bibliographystyle{plain}
}
\end{document}